# Field exposed water in a nanopore: liquid or vapour?

Dusan Bratko[a,b*], Christopher D. Daub[a] and Alenka Luzar[a*]

[a]Department of Chemistry, Virginia Commonwealth University, Richmond, VA 23284-2006, aluzar@vcu.edu
[b]Department of Chemical Engineering, University of California, Berkeley, CA 94720-1462, dnb@berkeley.edu

**Field-exposed water behaviour in hydrophobic confinement confirms classical electrostriction in nanoscale channels and nanoporous materials.**

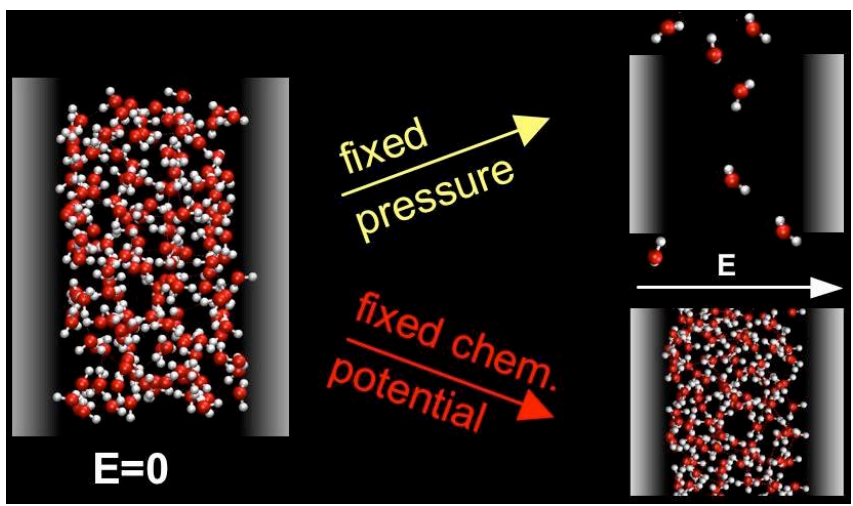

**Abstract**

We study the behavior of ambient temperature water under the combined effects of nanoscale confinement and applied electric field. Using molecular simulations we analyze the thermodynamic causes of field-induced expansion at some, and contraction at other conditions. Repulsion among parallel water dipoles and mild weakening of interactions between partially aligned water molecules prove sufficient to destabilize the aqueous liquid phase in isobaric systems in which all water molecules are permanently exposed to a uniform electric field. At the same time, simulations reveal comparatively weak field-induced perturbations of water structure upheld by flexible hydrogen bonding. In open systems with fixed chemical potential, these perturbations do not suffice to offset attraction of water into the field; additional water is typically driven from unperturbed bulk phase to the field-exposed region. In contrast to recent theoretical predictions in the literature, our analysis and simulations confirm that classical electrostriction characterizes usual electrowetting behavior in nanoscale channels and nanoporous materials.

**1. Introduction**

Because of their high dipole and quadrupole moments, water molecules feature strong interactions with electrostatic fields next to charged or polar solutes, and are attracted to field-exposed regions in electrowetting experiments[1]. Attractive interactions of water with electric field imply partial alignment of molecular dipoles with the direction of the field. For freely rotating molecules, the competition between the energy reduction and the loss of orientational entropy upon alignment is described by the well known Langevin equation. However, in liquid water, orientations of water molecules are also subject to angle restrictions associated with hydrogen bonding. In case of aqueous confinements, additional angular preferences are imposed by water's strong tendency to minimize the loss of hydrogen bonds at the interface[2-5]. Hydrogen bonding between water molecules favors near-parallel dipole orientations relative to the interface[3, 6-8], an arrangement that is also compatible with the general preference of *parallel* dipoles for vertical as opposed to horizontal coordination[9-11]. The structural behavior of water under combined effects of applied field and confinement is therefore quite complex and can ultimately depend on the choice of fixed system variables such as the number of particles, N, or chemical potential, µ. We pay particular attention to two situations addressed in recent theoretical studies[12-17]:

a) Confined or bulk mass-conserving isobaric system under uniform external field $E_o=|\boldsymbol{E_o}|$. Here, the number of bulk molecules $N$ in the field is fixed as are pressure and temperature $(N,p,T)$. Volume $V$ fluctuates around an average $<V>$, which may depend on the strength of the applied electric field $E_o$. We describe this variation by

$$\frac{\partial <V>}{\partial E_o} = \frac{\partial}{\partial E_o}(\Delta^{-1} \sum_{V_j} \sum_{states\ i} V_j e^{-\frac{U_i}{k_BT}-\frac{pV_j}{k_BT}}) = -(k_BT)^{-1}(<V\frac{\partial U}{\partial E_o}> - <V><\frac{\partial U}{\partial E_o}>)$$

$$\text{where}\quad \Delta = \sum_{V_j}\sum_i e^{-\frac{U_i}{k_BT}-\frac{pV_j}{k_BT}} \qquad (1)$$

Each of states *i* corresponds to a distinct configuration $[\boldsymbol{r}_N,\boldsymbol{\Omega}_N]$ consisting of positions *r* and orientations $\boldsymbol{\Omega}$ of all $N$ particles and the angle brackets denote the ensemble average. The slope $\partial U / \partial E_o$, related to the ease with which the molecules align with the field, is expected to increase with fluid dilution. Here, all *N* molecules are already exposed to the field and any structural rearrangement takes place only to find the best compromise between molecular alignment with the field and orientation-dependent interactions among molecules. The density of an isobaric polar fluid is therefore expected to decrease with the strength of the applied field $E_o$.

Classical electrowetting experiment, however, typically involve transfer of water to a field-exposed region to maximize direct water-field interaction[18]. Pressure and density are therefore not fixed in the second situation (b) we consider:

b) Isochoric aqueous confinement exposed to applied field $\boldsymbol{E_o}$ and open to the exchange of molecules with surrounding field-free ($|\boldsymbol{E_o}|=0$) reservoir of either liquid water or, equivalently, a water vapor phase at vapor pressure corresponding to given temperature $T$. The system is described by grand canonical ($\mu,V,T$) statistics with fixed volume ($V$), temperature, and chemical potential ($\mu$). The average number of molecules is given by

$$<N> \;=\; \Xi^{-1} \sum_N \sum_{\text{states } i} N e^{-\frac{U_i}{k_BT}+\frac{\mu N}{k_BT}} \quad \text{with} \quad \Xi = \sum_N \sum_i e^{-\frac{U_i}{k_BT}+\frac{\mu N}{k_BT}} \tag{2}$$

Electric field $\boldsymbol{E}$ affects $<N>$ through orientation-dependent interaction with molecular dipoles $\boldsymbol{\mu}_i$ reflected in potential energies $U_i$, leading to:

$$\partial N / \partial E_o = \frac{\mu}{k_BT}\left(<N^2 \frac{\partial (E_o \overline{\cos\theta_{N,i}})}{\partial E_o}> - <N><N \frac{\partial (E_o \overline{\cos\theta_{N,i}})}{\partial E_o}>\right) =$$

$$= \frac{\mu}{k_BT}\left(<N^2 \overline{\cos\theta_{N,i}}> - <N><N\overline{\cos\theta_{N,i}}>\right)$$

(3)

where $\mu=|\boldsymbol{\mu}|$. $\overline{\cos\theta_{N,i}}$ is the value of $\boldsymbol{\mu E_o}/|\boldsymbol{\mu E_o}|$ averaged over all $N$ molecules in the system at a specified configuration $i$. Because $\overline{\cos\theta_{N,i}} \geq 0$ for all representative configurations $i$, the *product* $(N\overline{\cos\theta_{N,i}})$ is usually a monotonically increasing function of $N$ although $\overline{\cos\theta_{N,i}}$ alone can, at certain conditions, be negatively correlated with the density. The density of a dipolar liquid in an open system is therefore generally expected to rise with increasing field strength $E_o$ as is predicted by continuum analyses[15, 19, 20] and seen in electrostriction experiments.

The opposite trend, $(\partial N / \partial E_o)_{\mu VT} < 0$, would only be possible in case of a dramatic rise in orientational polarizability of the molecules upon dilution, an idea explored in a recent mean-field analysis[17] of a polar fluid's phase behavior in electric field. In that study[17], an Ising model of a water-like fluid was used to consider a liquid-vapor phase transition in a system where intermolecular attractions were deemed incompatible with molecular dipole alignment with the field. Over an interval of intermediate field strengths, the system featured a density drop akin to field-induced capillary evaporation reported in an earlier simulation of confined water[15]. The majority of simulation studies of field-exposed confined water, including new open ensemble simulations in this work, however, conform to classical electrostriction behavior with water populations increased in the field[12, 14, 16, 21, 22]. These repeated observations lead us to question the hypothesis[17] of strongly negative correlation between attractive water-water interactions (dominated by hydrogen bonding), and water's ability to align with the applied field.

In the present study we explore the issue by Monte Carlo and Molecular Dynamics simulations whereby we directly monitor water structure and the extent of hydrogen bonding as the molecules become increasingly aligned by the applied field. We consider the range of *applied* field strengths $E_o = E\varepsilon_r$ up to 1 VÅ$^{-1}$. Due to water polarization, the *actual* field strength $E \sim E_o \varepsilon_r^{-1}$ remains below 0.1 VÅ$^{-1}$, spanning the range of fields detectable near charged electrodes, ion channels, ionic biomolecules or assemblies[21, 23-32]. Here and throughout the paper, $E_o = E\varepsilon_r$ stands for the unscreened applied field that does not include the field reduction due to water polarization $\rho\mu \overline{<\cos\theta_{N,i}>}$. $\varepsilon_r$ is relative permittivity and $\rho$ water number density. Within the above range of fields and concomitant dipole-field alignment $\overline{<\cos\theta_{N,i}>}$ up to ~0.8, we do not observe any significant field-induced changes in water hydrogen bond populations or in atom-atom distribution functions in the bulk phase. In narrow confinements, we observe only slight changes limited to the first solvation layer next to confinement walls. By-and-large, our calculations confirm the considerable resilience of the hydrogen-bond network under aligning electric field as reported in previous studies[27, 29, 33, 34]. This behavior is rationalized by the flexibility of hydrogen-bond angles. For ambient temperature (~300 K), conventional hydrogen-bond definitions[35] envisage about 30 degree tolerance from the zero-temperature bond angle. Calculations we describe below demonstrate this flexibility suffices to accommodate relatively high alignment of aqueous dipoles without serious penalties in the number and free energies of hydrogen bonds.

## 2. Density changes of bulk and confined water under electric field

To establish a common reference, we first performed molecular simulations analogous to those of England *et al*[17], who studied the density of field-exposed bulk water at fixed ($N,p,T$) using molecular dynamics with a 1 nm cutoff of intermolecular interactions. The choice of truncation is motivated by the interest in the behavior of confined water where the omission of long-ranged forces mimics confinement effects on water-water correlations[17]. We use Extended Simple Point Charge model of water[36] (SPC/E) employed in previous related studies[14-16]. Fields we consider, $E=O(10^{-2}-10^{-1})$V Å$^{-1}$ ($E_o \leq 1$ V Å$^{-1}$), are weak compared to those around simple ions and do not warrant[37] the use of polarizable models of water[38, 39]. In molecular dynamics simulations we performed with the DL_POLY package[40], the average pressure was set to the ambient pressure by a Berendsen barostat referring to the average of pressure tensor components. In simultaneous Monte Carlo calculations, interpolation from a series of ($NVT$) calculations was used to identify densities supporting the original pressure after the introduction of the field. Identical calculations were then repeated in nanosized planar confinements[41] with the wall/water interaction described by the (9-3) integrated Lennard-Jones potential[3, 6, 42, 43]. Our model system has been described in detail elsewhere[16, 44, 45]. In summary, it consists of a thin slab of water confined between two continuous apolar plates. The system is periodic in the *x* and *y* directions. In most calculations, we used smooth 1 nm spherical truncation of intermolecular potential[6]. In a subset of calculations, these results were validated by using the slab-adapted Ewald sum method described in ref. [46]. The slab of water is considered to be in equililbrium with an outside reservoir of bulk water. We impose this condition by computing the chemical potential $\mu$ required to produce atmospheric pressure in the bulk system. When a field is applied across the plates, the system's chemical potential remains unchanged; in other words, the bulk reservoir is not considered to be exposed to the field.

We considered three different strengths of wall/water Lennard-Jones energy parameter, associated with water contact angles $\theta_c$=135, 93, and 69°. In a separate study we show[8] these angles correspond to water-wall site Lennard-Jones coupling parameters $\varepsilon$=0.648, 3.45 and 5.0 kJ mol$^{-1}$, respectively. Distance $D$ was set at 16.4 Å, the smallest separation that still avoids capillary evaporation[42, 45, 47, 48] at the highest of the three contact angles considered. In Fig. 1, we present water density relative to that of a field-free system ($E_o$=0) as a function of field strength $E_o$. We compare effects of applied field on two systems, bulk and confined, in each system monitoring the density change upon introduction of the field under two different, (fixed chemical potential or pressure) scenarios. In agreement with the mean-field prediction[17] and our present analysis (Eq. 1), attractions among a fixed number of particles are loosened upon introduction of competing field-induced molecular reorientation. The critical temperature of water is hence lowered in electric field. At fixed pressure, increasing applied field results in gradual expansion and eventual liquid/gas phase separation. Such evaporation was actually observed in ref.[17] and possibly in ref.[15,49]. Our (*NpT*) Molecular Dynamics calculations display a similar density depression resembling the onset of evaporation but we do not observe an actual phase transition in the course of up to ns simulation runs. Barriers to homogeneous vapor nucleation may have played a role. Activation barrier issues were avoided in our Monte Carlo calculations with preset densities. Drawing a curve connecting densities with fixed pressure (or selected pressure tensor component) at different field strengths, our Monte Carlo calculations show a trend toward expansion *and* eventual liquid-to-vapor transition in both, bulk and confined systems at isobaric (*NpT*) conditions.

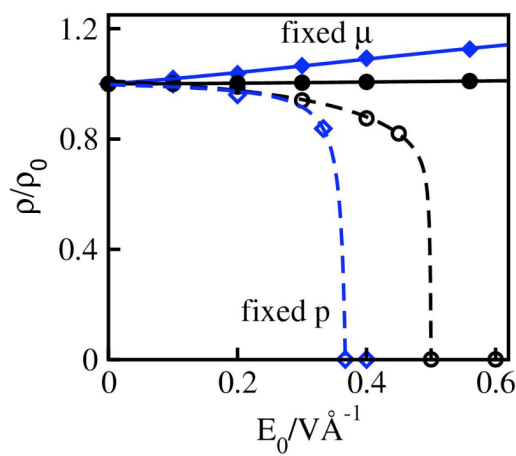

**Fig. 1** Effect of applied electric field $E_0$ on density of SPC/E water (relative to that observed in the absence of applied field) in bulk phase (circles) and in a planar confinement of width D=16.4 Å and wall/water contact angle $\theta_c$=135° (diamonds). Solid symbols and lines correspond to open systems with fixed chemical potential ($\mu VT$) and open symbols represent points with fixed pressure ($NpT$) for $p_{zz}(E_0) = p_{zz}(0) \pm 20$ atm. In the confinement, the field points in $z$ direction (normal to the walls). Temperature is 298 K.

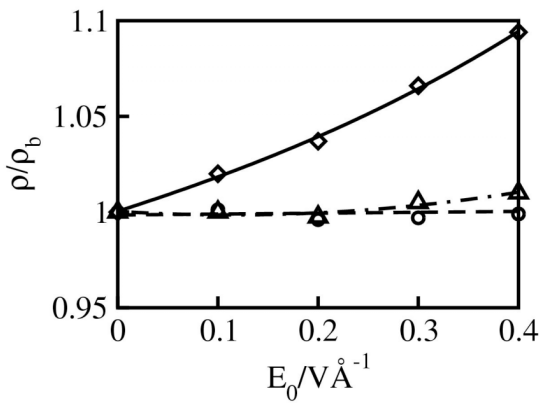

**Fig. 2** Electrostriction of water in *open* (*μVT*) nanoscale confinements (D=16.4 Å) with walls of different hydrophobicities/philicities quantified in terms of contact angle: $\theta_c$=135° (diamonds), 93° (triangles), and 69° (circles).

In a system with constant chemical potential where additional water is allowed to enter in response to field exposure, our calculations reveal consistent increase in water density in the field, both in the bulk and confined cases. While the change is relatively small in poorly compressible bulk water, we observe a considerable density rise in strongly hydrophobic confinements. In the latter case, compression is primarily a result of filling up depleted surface layers next to hydrophobic walls[50, 51]. In Fig. 2, we compare the dependencies of the density on the field strength for water confined in pores of width 16.4 Å for three different contact angles, $\theta_c$=135°, 93°, and 69°. Only for strongly hydrophobic walls ($\theta_c$=135°) we observe significant electrostriction. The density increases only slightly in pores with intermediate or hydrophilic walls. This behavior conforms with calculated compressibility dependence on the applied field for the three cases shown in Fig. 3. In the absence of the field, water between strongly hydrophobic plates appears much more compressible than in the bulk phase[42, 52]. This effect is absent at lower contact angles ($\theta_c$=93° or 69°) where little or no surface density depletion takes place. Applied field increases the apparent hydrophilicity. Upon increasing the field strength, compressibilities in all three systems asymptotically converge toward that of the bulk phase.

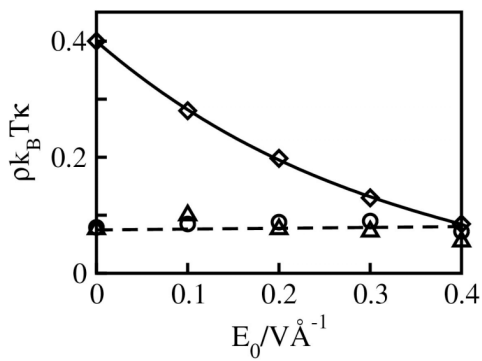

**Fig. 3** Reduced compressibilities in aqueous confinements of width $D$=16.4 Å with walls of different contact angles: $\theta_c$=135° (diamonds), 93° (triangles), and 69° (circles) in the presence of applied electric field $E_0$.

The difference between the observed behaviors of systems at constant-pressure (bottom portion of Fig. 1) and constant chemical potential (upper portion in Fig. 1) persisted in additional calculations for stronger fields of up to several V Å$^{-1}$ (not shown). In open systems, the weakening of intermolecular water-water attraction upon molecular alignment could not offset the free-energetic incentive to drive the molecules into the field-exposed region. To substantiate this assertion, we monitored explicitly the contributions to molecular energies due to dipole-field interaction, $(\mu E \overline{\cos\theta_{N,i}})$, and compared them with intermolecular interactions and specifically with interaction due to hydrogen bonding.

**3. Interactions among water molecules aligned by applied electric field**

In Table I, we list calculated energies and average numbers of hydrogen bonds per water molecule in bulk water, all as functions of applied field strength. For the analysis of water hydrogen bonding we use our usual definition based on geometric criteria[35]: two water molecules can be either bonded or not bonded depending upon their distance between an oxygen atom acting as proton acceptor and a hydrogen of the molecule whose oxygen atom acts as a proton donor, i. e. OH intermolecular distance, and the angle between the O-O vector and the covalent O-H bond. The numerical cutoff value we use for $r_{OH}^c$ = 2.45 Å corresponds to the first minimum in the SPC/E water radial distribution function, $g_{OH}(r)$. The value of the cutoff angle $\alpha^c = 30^0$ is the angle at which the average number of hydrogen bonds per water molecule is within 10% of the asymptotic value for large $\alpha^c$ [35, 53]. Interestingly, our calculations under electric field show this threshold value not to change up to field strengths of 1.5 VÅ$^{-1}$ revealing the remarkable plasticity of the water's hydrogen bond network. In a preceding study[54], we presented a comparison between field effects on average numbers of hydrogen bonds determined by both the geometric[35] and energetic[55] criteria. Identical effects were found using either method, hence only the geometric definition is applied here.

**Table I.** Average energy per water molecule, interaction of a molecular dipole with the applied field, ($U_f$), number of hydrogen bonds per water molecule (either for the whole system, or only for waters in interfacial regions, $\pm z$), Coulombic ($U_{HB}^c$) and total ($U_{HB}^t$) interaction between a pair of hydrogen-bonded molecules, and the average angle between molecular dipoles and the direction of the field, $\overline{\cos\theta_{N,i}}$. Statistical error bars are below $\pm 1$ at the last digit reported.

*Bulk phase*

| $\dfrac{E_o}{V\text{Å}^{-1}}$ | $\dfrac{<U>}{Nk_BT}$ | $\dfrac{<U_f>}{Nk_BT}$ | $<n_{HB}>$ | $\dfrac{<U_{HB}^C>}{Nk_BT}$ | $\dfrac{<U_{HB}^t>}{Nk_BT}$ | $\overline{\cos\theta_{N,i}}$ |
|---|---|---|---|---|---|---|
| 0 | -19.3 | 0 | 3.53 | -10.6 | -7.8 | 0 |
| 0.2 | -19.5 | -0.73 | 3.53 | -10.4 | -7.6 | 0.19 |
| 0.4 | -20.6 | -2.9 | 3.53 | -10.4 | -7.7 | 0.37 |
| 0.6 | -22.2 | -6.3 | 3.54 | -10.2 | -7.4 | 0.55 |
| 0.8 | -24.9 | -10.8 | 3.54 | -9.8 | -7.0 | 0.71 |
| 1.0 | -28.2 | -15.4 | 3.57 | -9.6 | -6.8 | 0.80 |

*Confinement* ($D$=16.4 Å)

| $\dfrac{E_o}{V\text{Å}^{-1}}$ | $\dfrac{<U>}{Nk_BT}$ | $\dfrac{<U_f>}{Nk_BT}$ | $<n_{HB}>$, all | $<n_{HB}>$, $-z$ | $<n_{HB}>$, $+z$ | $\dfrac{<U_{HB}^C>}{Nk_BT}$ | $\dfrac{<U_{HB}^t>}{Nk_BT}$ | $\overline{\cos\theta_{N,i}}$ |
|---|---|---|---|---|---|---|---|---|
| 0 | -18.3 | 0 | 3.29 | 2.57 | 2.57 | -10.6 | -7.8 | 0 |
| 0.2 | -18.6 | -0.58 | 3.31 | 2.80 | 2.52 | -10.7 | -7.9 | 0.153 |
| 0.4 | -19.6 | -2.2 | 3.34 | 2.95 | 2.58 | -10.5 | -7.7 | 0.29 |
| 0.6 | -20.9 | -4.6 | 3.33 | 2.96 | 2.60 | -10.4 | -7.7 | 0.407 |
| 0.8 | -23.1 | -8.3 | 3.30 | 2.89 | 2.63 | -10.2 | -7.4 | 0.55 |

*Confinement* ($D$=16.4 Å) with 3D slab-corrected Ewald sums[46].

| $\dfrac{E_o}{V\text{Å}^{-1}}$ | $<N>$ | $\dfrac{<U>}{Nk_BT}$ | $\dfrac{<U_f>}{Nk_BT}$ | $<n_{HB}>$ | $\dfrac{<U_{HB}^C>}{Nk_BT}$ | $\dfrac{<U_{HB}^t>}{Nk_BT}$ | $\overline{\cos\theta_{N,i}}$ |
|---|---|---|---|---|---|---|---|
| 0 | 168.8 | -17.2 | 0 | 3.26 | -10.6 | -8.0 | 0 |
| 0.2 | 170.1 | -17.6 | -0.37 | 3.28 | -10.7 | -8.0 | 0.096 |
| 0.4 | 174.9 | -18.7 | -1.40 | 3.29 | -10.6 | -7.9 | 0.183 |
| 0.6 | 180.0 | -19.7 | -2.99 | 3.31 | -10.6 | -7.8 | 0.261 |
| 0.8 | 185.8 | -20.8 | -4.87 | 3.32 | -10.6 | -7.8 | 0.32 |
| 1.0 | 189.5 | -22.1 | -7.26 | 3.32 | -10.5 | -7.7 | 0.38 |

Our calculations reveal a strong overall reduction (increase in absolute value) of molecular energies upon increasing the field strength $E_o$. The energy reduction is weaker than the average dipole interaction with the applied field, $U_f$, primarily due to the mitigating effect of water polarization opposing the applied field. We also include calculated Coulombic and total interaction energies for pairs of H-bonded molecules. These data show that any change in these energies arises primarily from the Coulombic part.

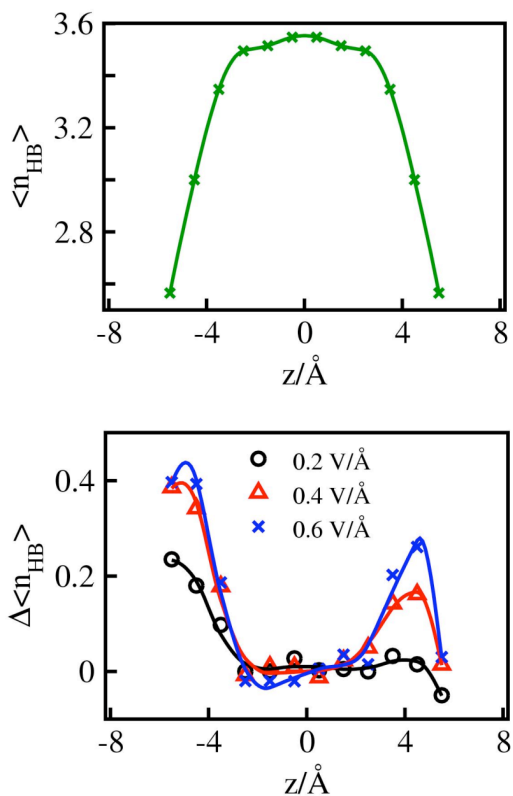

**Fig. 4** The average number of hydrogen bonds $<n_{HB}>$ maintained by a water molecule in field-free system (top), and the electric field induced *change* in the number of hydrogen bonds, $\Delta<n_{HB}>=<n_{HB}(E)-n_{HB}(0)>$ (bottom) in a 16.4 Å wide hydrophobic confinement as function of molecular position $z$ for fields of strength $E_0$=0.2, 0.4, and 0.6 V Å$^{-1}$.

For reported field strengths of up to 1 V Å$^{-1}$, as well as for much higher fields (not shown), the prevalence and strength of hydrogen bonding are only weakly related to molecular alignment (measured in terms of $\overline{cos\theta_{N,i}}$ ). In confined systems, the average number of bonds per molecule can actually increase upon application of the field, the difference being noticeable especially in interfacial layers next to confinement walls, Fig. 4. Because of water asymmetry, the change depends on the direction of the field relative to the surfaces, leading to a notable asymmetry in hydrogen bond density profiles across the confinement. Concomitant asymmetries in density profiles have been pointed out by a number of groups[14-16].

Small changes in hydrogen-bond populations upon molecular alignment with the field conform with the observed insensitivity of atom-atom (O-O, O-H and H-H) radial distribution functions $g(r)$ to the field in the range of our interest ($E_o \leq 1$ V Å$^{-1}$). For these fields, the changes of bulk $g(r)$-s are insignificant and are hence not shown. (An onset of the transition from tetrahedral toward tightly-packed, highly coordinated "electrofrozen" structure[27, 56, 57] can be observed at stronger fields of about 2 V Å$^{-1}$ and higher).

More subtle changes in hydrogen bond network, including any deviations from tetrahedrality might be reflected in the distributions of the O-O-O angles in triplets of nearest-neighbor molecules. These distributions measure the plasticity of the network and characterize fluctuations around the apparent tetrahedral coordination[58, 59]. However, our results, collected in Fig. 5, show no significant changes in (O-O-O) angle distributions either.

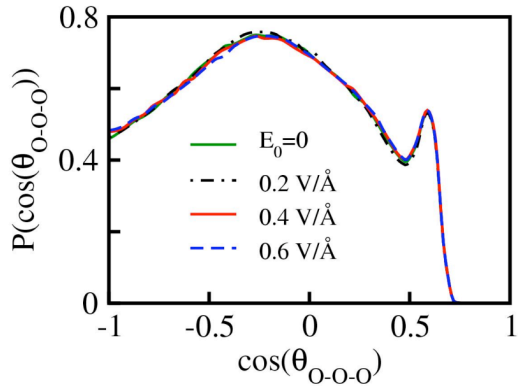

**Fig. 5** Probability distributions of oxygen triplet angle at different field strengths. Triplets are enumerated out of all nearest neighbours ($r_{OO} < 3.4$ Å) of each water molecule, where the central water always forms the apex of the angle.

Under additional rotational restrictions in the field, small differences in hydrogen bond energies shown in Table I could still be compatible with more significant changes in corresponding free energies. If so, however, the weakening of hydrogen-bond free energies would reflect in reduced bond populations. We do not observe such reductions in our simulations. Despite mild weakening of water-water interactions, in open ensemble cases, this trend is more than offset by the attraction of dipolar water molecules into the field region. From Eq. (3), applied to high fields $E_o$, it is obvious that the density will increase monotonically when dipoles reach the strong-alignment limit. Physically, this is explained by the fact that the repulsion between aligned dipoles reaches a plateau value whereas the dipole-field energies continue to fall in proportion to field strength $E_o$. In an open system, electrostriction therefore represents the general behavior; if density depression would occur, it would necessarily feature a transient dependence on $E_o$ as predicted in Fig. 2 of ref.[17]. Our present calculations, as well as simulation studies by other groups[27, 29, 33, 60], however, reveal relatively mild weakening of hydrogen bonds due to the application of the field. This weakening may suffice to trigger a liquid to vapour transition of water at constant pressure and fixed number of molecules in the field, but not in an open system where evaporation would entail transfer of water from the field region to field-free surroundings.

Simulation at constant chemical potential appears the natural choice to mimic small pore electrowetting. This can be readily implemented in (Grand Canonical) Monte Carlo simulations[12, 16]. An alternative setup, particularly suitable in Molecular Dynamics studies involves a simulation of both the confinement and bulk, field-free environment (reservoir) in equilibrium with each other[14, 15]. Maintaining constant pressure in the bulk phase will fix the chemical potential. However, the use of barostat algorithms requires additional care as pressure can be strongly nonuniform, anisotropic, and even discontinuous in case of discontinuous[15] electric field. The latter problem is avoided if the field stems from internal charges or is introduced as a smooth function of position.

## 4. Effect of potential cutoff and boundary conditions

Finally we mention the comparatively strong dependence of simulated water polarization by the field on potential cutoff and applied boundary conditions. This dependence has to be kept in mind in discussing the field effect on water behavior in context of molecular simulation. In Fig. 6, we collect the data for the average molecular alignment (quantified in terms of $<\cos\theta>$) as a function of field strength $E_o$ for different conditions. To illustrate the issue, we include data from refs.[14, 15, 27], and over a range of fields much broader than considered in the bulk of this work. We note the sensitivity of the polarization response of the water dipoles to the use of different boundary conditions (vacuum *vs* tin-foil) in the results of Ewald sum calculations when electric field is applied. With conducting ("tin foil"[61]) boundaries, the favorable energy gained by aligned dipoles causes a much stronger response to the field compared with calculations with a spherical potential cutoff. Conversely, the use of vacuum boundaries introduces a surface correction term[61] which accounts for the presence of a depolarized surface in contact with surrounding vacuum. In confinement, this leads to weaker polarization than seen in the cutoff calculation. For our particular choice of the cutoff distance and system size, polarization of bulk water observed using spherical cutoff was similar to that obtained from Ewald sum calculation with the vacuum boundary condition. The choice of boundary terms must be motivated by careful consideration of the experimental conditions the simulation is meant to mimic. We find the vacuum-boundary correction to be best suited to Ewald sum studies of field-exposed bulk water, and related slab-corrected Ewald sums for the confined water system[46] to describing field-enhanced wetting of nonpolar nanopores.

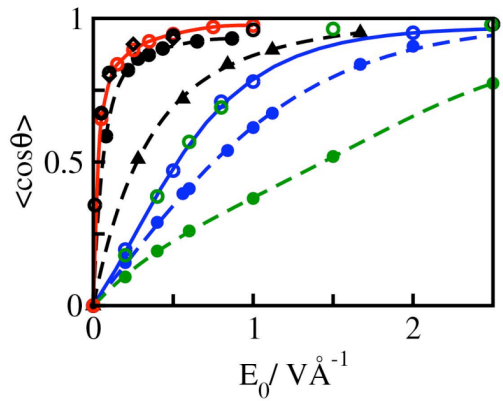

**Fig. 6** Simulated average alignment of water dipoles in bulk (open symbols) and confined (solid symbols) SPC/E water. Bulk data are from Ewald sum calculations carried out in this work, both with conducting (red) and vacuum (green) boundary conditions, and by using 10 Å spherical truncation of water-water potentials (open blue circles). Solid blue and green symbols represent respectively cutoff and slab corrected 3-dimensional Ewald sum[46] results for planar, 16.4 Å wide confinement, both from this work. The black symbols and lines describe Ewald sum results for electric field exposed water from other labs: bulk water with conducting boundaries from ref.[27] (open circles) and ref.[14] (open diamonds; note that the applied field -- our $E_0$ -- in bulk simulations is $<E>$ in ref. 14, while their $<E_0>$ includes an effective screening contribution), confined water in a 16.4 Å wide rectangular confinement surrounded by field-free reservoir from ref.[15] (solid circles), or in cylindrical, 20 Å wide ion channels, surrounded partially by channel walls and open to the bulk phase, ref.[14] (solid triangles).

However, because of the high dielectric constant of water, at moderate fields confined systems surrounded by field-free water reservoirs (solid black symbols in Fig. 6) better resemble situations subject to tin-foil boundary conditions.

At otherwise identical conditions, introduction of Ewald sums generally results in stronger dielectric screening of applied field and hence in weaker orientational polarization (lower $<\cos\theta>$; See the last column in Table I). This, in turn, means weaker repulsion of parallel molecular dipoles[62]. The difference conforms with the observed phase behavior of England *et al*[17]. They show that at constant pressure conditions (*NpT*), bulk TIP4P water, modeled using pair potential truncated at 1 nm expands under electric field, but contracts when cutoff is removed and Ewald sums are used[17]. We observe the same trends in isobaric simulations with the SPC/E model of water. As already pointed out[17], potential truncation mimics conditions representative of water under confinement and Ewald summation pertains to an extended bulk phase.

A qualitative implication of the above comparisons, relevant to our calculations, is that truncation of pair potentials is comparatively more conducive to field-induced isobaric expansion and possibly evaporation. The consistent electrostriction we observe in open ensemble ($\mu VT$) calculations therefore cannot be attributed to the use of potential cutoff. To remove any uncertainty, we verify this assertion in direct slab-corrected 3-dimensional Ewald sum[46] calculations for our confined systems using GCMC simulation. Average numbers of molecules, $<N>$, observed in confinement runs listed in the bottom part of Table I, and represented by solid green symbols in Fig. 6, as a function of applied field strength $E_o$ confirm monotonic electrostriction behavior for ($\mu VT$) conditions, independent of the use of either potential truncation or Ewald summation.

## 5. Concluding remarks

In conclusion, we revisit the problem of simulating confined water in an electric field, stressing the importance of open ensembles, and the ability of water to maintain hydrogen bonding despite molecular orientation in the field. We find water hydrogen-bond interactions surprisingly compatible with partial molecular alignment under applied electric field. As a result, water molecules in bulk and confined phases alike, can sustain much of their mutual attractions while the field simultaneously polarizes them. At constant pressure conditions, weakening of intermolecular attractions can result in expansion of the fluid and possible liquid-to-vapor transition, consistent with the mean-field predictions of Pande and coworkers[17]. When field-exposed nano-sized confinement is equilibrated with bulk, field-free phase with fixed chemical potential, however, the present analysis suggests water will be attracted to the confinement to increase exposure to electric field, typically resulting in conventional electrostriction behavior.

## Acknowledgments

We thank Max Berkowitz and Jay Rasaiah for helpful discussions preceding this study, Joachim Dzubiella for help interpreting results of Ref. 14, and National Science Foundation for support through awards CHE-0718724 (to A.L.) and CBET- 0432625 (D.B.).